\documentclass[aps,pra,twocolumn,floatfix,amsmath,amssymb,]{revtex4-2}
\usepackage{physics}
\usepackage{graphicx}
\usepackage{bbm}
\newcommand{\cD}{\mathcal{D}}
\newcommand{\cL}{\mathcal{L}}
\newcommand{\cK}{\mathcal{K}}
\newcommand{\cA}{\mathcal{A}}
\begin{document}
\title{Dressed-state master equation for two strongly coupled two-level atoms with long-lived entanglement} 
\author{Artemisa Villalobos-Ramirez}
\author{Juan Mauricio Torres}
\email{jmtorres@ifuap.buap.mx}
\affiliation{Instituto de F\'isica, Benem\'erita Universidad Aut\'onoma de Puebla, 72570 Puebla, M\'exico }
\date{\today}

\begin{abstract}
We derive a dressed-state master equation in Lindblad form for two strongly coupled two-level atoms. The resulting decay dynamics are governed by Lindblad operators that couple different dressed states. We show that the eigenvalues and eigenvectors of the Liouvillian can be obtained in a compact form, since each off-diagonal element in the dressed-state basis constitutes an eigenvector. Depending on the interatomic distance and the atomic transition frequency, the decay exhibits two well-separated time scales. On short times, the system relaxes into a pair of states, one of which is a transient, maximally entangled state. On longer times, this intermediate entanglement irreversibly decays into a separable steady state. Our results demonstrate that the intrinsic decay mechanism can transiently generate maximal entanglement, an effect that is not captured without the dressed-state master-equation formalism.
\end{abstract}

\maketitle

\section{Introduction}

The theory of open quantum systems has been remarkably successful in describing quantum systems in close correspondence with experimental reality, many of which now form the basis of novel quantum technological applications \cite{Carmichael1993,BreuerPetruccione2002,manzano2020,Ashida2020}. Within the Markovian regime, the Gorini–Kossakowski–Sudarshan–Lindblad (GKSL) master equation provides the most general form of a dynamical generator for completely positive, trace-preserving semigroups, and serves as the standard starting point for analyzing dissipation and decoherence \cite{Rivas2012,Xia2024,Long2024,YanesThomas2026}. A wide range of experimental architectures, from cavity and circuit QED to trapped ions, neutral atoms, and semiconductor or superconducting qubits, can be accurately modeled within this framework \cite{Haroche2006,Blais2021,Wendin2017,Noiri2022,Evered2023,Lu2019,Akram2022}.

For composite interacting systems, it has long been common practice to employ a phenomenological master equation constructed by simply adding the interaction to the Lindblad master equations of the separate subsystems. This approach has endured over the years, as it offers a good description when the interaction within the central system is weak \cite{Rivas2012,Xia2024}. In contrast, recent advances offer the possibility of achieving strong internal coupling \cite{Browaeys2016,vanDitzhuijzen2008, Du2023, Chomaz23}. This is not only a scientific curiosity, but it also enables improved control and manipulation of key features in composite systems. At the same time, it highlights the limitations of the phenomenological construction, which can fail to capture important aspects of the dissipative dynamics, including steady-state properties and entanglement generation \cite{Vaaranta2024,Smith2021,Soares2022}. Therefore, understanding dissipative dynamics in strongly coupled quantum systems is crucial for quantum information processing \cite{Haroche2006,Blais2021,Wendin2017,Noiri2022,Evered2023,Lu2019,Akram2022,Lu2012}.

The dressed-state master equation provides an alternative approach to describing such strongly interacting systems by deriving the dynamics from first principles using the eigensystem of the central system, or dressed states \cite{Hu2014}. This framework has gained attention in the field of optomechanics \cite{Hu2014,Naseem2018,Torres2019,Betzholz2020}. Another prominent example is the Jaynes–Cummings model \cite{Scala2007}, where the resulting Lindblad operators describing relaxation are expressed in terms of the dressed states of the atom–mode Hamiltonian. Since dressed states are typically entangled, the resulting decay transitions operate between these entangled states, raising the question of whether the dissipative dynamics can be exploited to preserve or even generate entanglement in the system.

Here, we address this question by deriving a dressed-state master equation for two strongly dipole-coupled two-level atoms, relying on the standard formalism \cite{BreuerPetruccione2002,Scala2007}. We take into account the atomic separation and the spatial dependence of the electromagnetic field at the different atomic positions, and we identify the Lindblad operators that connect different dressed states of the central system. 

We show that, for certain parameter regimes, the dissipative dynamics generate a transient entangled state that survives on a time scale much longer than the primary relaxation process before eventually decaying into the steady state. This  behavior is  reminiscent of other protected or engineered subspaces in open systems \cite{Wu2021,Soares2022,AbdelWahab2023,Sadiek2021,Jahanbakhsh2022}, which can be highly appealing for applications in quantum technologies. These results reveal how decay processes between dressed states can transiently generate entanglement, a feature absent in the corresponding phenomenological description.

This work is organized as follows. Section \ref{sec:derivation} presents the model and the derivation of the dressed-state master equation assuming strong coupling between the atoms. We also discuss the parameter regime where the presented equation is valid and identify two separate time scales: a short one where a maximally entangled state is decoupled from the dynamics, and a second, long-time scale over which it eventually decays. In Section \ref{sec:zerotemp}, we completely solve the eigenvalue problem for the zero-temperature case and present the eigensystem in a compact form. We use this solution to derive a simple expression for the concurrence, analytically demonstrating the preservation of entanglement in a transient way. In Section \ref{sec:nonzerotemp}, we analyze the scenario of a non-zero temperature bath, where we find analytical expressions for the transient steady state and use them to calculate the concurrence, verifying our findings with numerical calculations. A comparison with the phenomenological master equation is presented in Section \ref{sec:Phenomenological}. Finally, Section \ref{sec:conclusions} summarizes the main findings and conclusions.
%%%%%%%%%%%%%%%%%%%%%%%%%%%%%%%%%%%%%%%%%%%%%%%%%%%%%%%%%%%%%%%%%%%%%%%%%%%%%%%%%%%%%%%%%%%%%%%%%%%%%%
\section{Derivation of the dressed-state master equation}\label{sec:derivation}
\subsection{The Hamiltonian model}
We consider an open quantum system composed of two two-level atoms, acting as the central system, coupled to an environment consisting of the surrounding electromagnetic field, which is described as a collection of harmonic oscillators. The total Hamiltonian describing the dynamics is given by
\begin{equation}
    H = H_S + H_E + H_{SE},
\end{equation}
where $H_S$ and $H_E$ generate the unitary dynamics of the central system and its environment, respectively, while $H_{SE}$ accounts for their interaction. The system Hamiltonian for the two identical two-level atoms can be written as
\begin{equation}\label{2atom}
H_S = \hbar \Omega \sum_{\mu=1}^2\sigma_\mu^+ \sigma_\mu^-  + \hbar g (\sigma_1^+ \sigma_2^- + \sigma_1^- \sigma_2^+),
\end{equation}
where $\mu\in \{1,2\}$ labels each of the two atoms, $\sigma_\mu^+$ ($\sigma_\mu^-$) denotes the raising (lowering) operator of the $\mu$-th atom, $\Omega$ is their transition frequency, and $g$ is the coherent coupling constant arising from a dipole-dipole interaction that will be discussed in more detail later. 
The environment consists of the quantized modes of the electromagnetic field, modeled as a collection of harmonic oscillators whose Hamiltonian is expressed as
\begin{equation}
    H_E =  \sum_{\vec k,\lambda} \hbar\omega_{k} a_{\vec k,\lambda}^{\dagger} a_{\vec k,\lambda},
\end{equation} 
where $\omega_k$ are the mode frequencies, $\lambda$ accounts for the polarization, $\vec k$ is the wave vector of each mode with $k=|\vec k|$, and $a_{\vec k,\lambda}$ are the annihilation operators for each mode. 

\begin{figure}
	\centering
	\includegraphics[width=.7\linewidth]{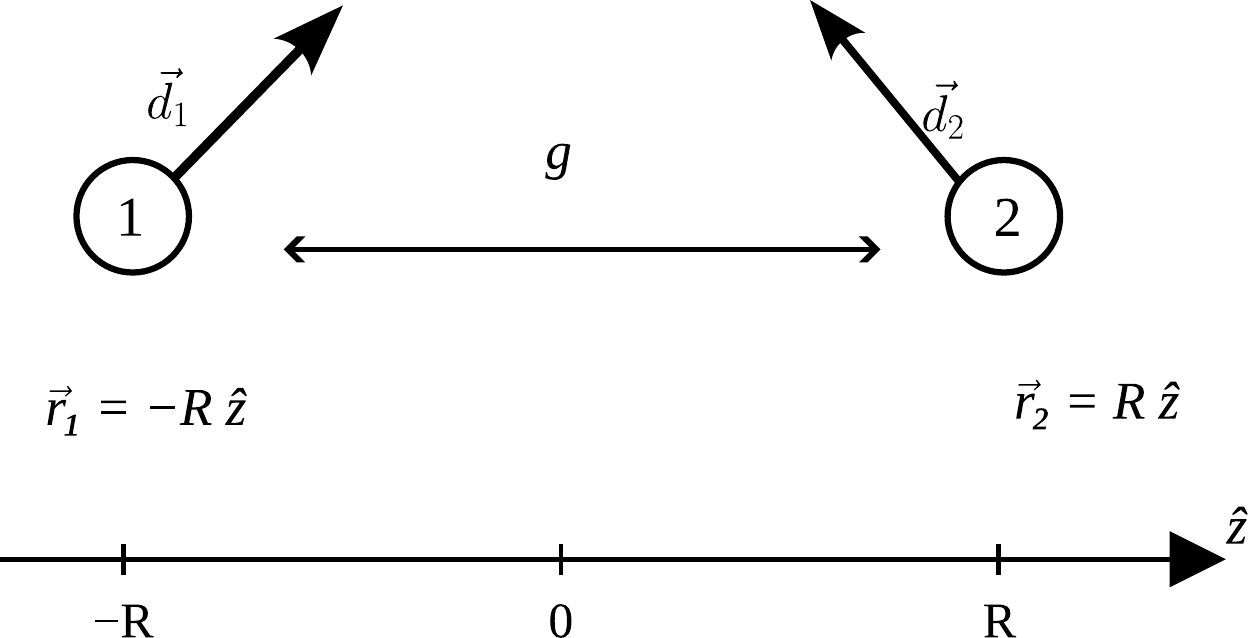}
	\caption{ Schematic of two atoms positioned symmetrically along the $z$-axis at $\vec r_1=-R\hat{ z}$ and $\vec r_2=+R\hat{z}$ . The dipole moments $\vec d_1$ and $\vec d_2$ can be oriented in any direction. The coupling strength between the atoms is denoted by $g$.}
	\label{fig:dipolesR}
\end{figure}

The interaction between the central system and its environment is described by the electric dipole Hamiltonian \cite{BreuerPetruccione2002,Carmichael1993}, which can be written in the following form:
\begin{equation}
\label{eq:int}
    H_{SE} = -\sum_{\mu=1}^2 \vec{D}_\mu \cdot \vec{E}(\vec{r}_\mu) = -\sum_{\mu=1}^2\sum_{l=1}^3
    D_{\mu,l}E_{\mu,l},
\end{equation}
where $\vec{E}(\vec{r}_\mu)$ is the electric field at the position of atom $\mu$, whose $l$-th Cartesian component, $l \in\{ 1,2,3\}$, is denoted by $E_{\mu,l}$. We have also introduced the atomic dipole-moment operator,
\begin{equation} \label{eq.dipolar.mu}
   \vec D_\mu =\vec d_\mu \sigma_\mu^- + \vec d_\mu^* \sigma_\mu^+ .
\end{equation} 
which is expressed in terms of the dipole vector $\vec d_\mu$. The atoms are assumed to be positioned symmetrically at $\vec r_2 = -\vec r_1 = R\hat z$, as depicted in Fig. \ref{fig:dipolesR}. With these considerations, the electric field at the positions of the atoms is given by $\vec E_1 \equiv \vec E(-R\hat z)$ and $\vec E_2 \equiv \vec E(R \hat z)$ \cite{Schleich2015}, and can be written as
\begin{equation}
	\label{eq:ElectricField}
    \vec{E}_\mu = \sum_{\vec{k}, \lambda} 
    \sqrt{\frac{\hbar \omega_k}{2 \epsilon_0 V}} 
    %\left( 
    a_{\vec{k}, \lambda}  e^{i(-1)^\mu  \vec{k} \cdot ( R \hat{z})}  %a_{\vec{k}, \lambda}^\dagger  e^{i(-1)^{\mu+1} \vec{k} \cdot \vec{R}} \right) 
    \hat{e}_{\vec{k}, \lambda}+ {\rm H.c.},
\end{equation}
where $\epsilon_0$ is the vacuum permittivity, $V$ is the quantization volume, and $\hat e_{\vec k, \lambda}$ is a unit polarization vector for each mode.

\subsection{Tracing over the field degrees of freedom}
In order to derive a Markovian master equation for the two atoms, one must trace over the environmental degrees of freedom. We will follow the procedure given in Ref. \cite{BreuerPetruccione2002} and summarize the results in this section, supported by detailed calculations in Appendix \ref{sec:derivemaster}. Certain peculiarities arise in our case, which will be further explained, primarily because the atomic separation causes the electromagnetic field to differ at each atom's position \cite{BreuerPetruccione2002}. 

To begin the derivation, the set of eigenvalues and eigenstates of the system Hamiltonian, $H_S$, is required. It is not a complicated task to verify that  the eigenstates of $H_S$  are given by
\begin{equation}\label{eq:dressedbasis}
\ket{0} = \ket{gg},\quad \ket{i} = \frac{\ket{ge} + (-1)^i \ket{eg}}{\sqrt{2}},\quad \ket{3} = \ket{ee}.
\end{equation}
Entanglement is a common feature of the eigenstates of interacting quantum systems. Indeed, the states $\ket{1}$ and $\ket{2}$ are maximally entangled. In contrast, $\ket{0}$ and $\ket{3}$ are completely separable; nevertheless, we will refer to this entire set as the dressed basis, as it constitutes the eigenbasis of the interacting system. The corresponding eigenenergies, $\hbar \Omega_j$, can be written in terms of the eigenfrequencies
\begin{equation}
	\label{eq:dressedOmegas}
  \Omega_0 = 0 , \quad \Omega_i  = \Omega + (-1)^i g,\quad  \Omega_3 = 2 \Omega,  
\end{equation}
considering $i\in\{1,2\}$. In this way, the diagonal form of the system Hamiltonian is given in the dressed basis as $H_S=\sum_{j}\hbar \Omega_j\ketbra{j}{j}$.

With the aid of the dressed-state projectors $\ketbra{j}{j}$, the components of the atomic dipole operator, $D_{\mu,l}$, can be decomposed into different frequency contributions in the following way
\begin{equation}
 D_{\mu,l}(\omega)\;\equiv\;\sum_{\Omega_{j'}-\Omega_j=\omega}
\ketbra{j}{j}\,D_{\mu,l}\,\ketbra{j'}{j'}, 
\end{equation}
where the sum is taken over every possible frequency difference such that $D_{\mu,l}=\sum_\omega D_{\mu,l}(\omega)$. There are four relevant values of all possible frequency differences, $\omega$, namely $\Omega_3-\Omega_2=\Omega_1-\Omega_0=\Omega_1$,  $\Omega_3-\Omega_1=\Omega_2-\Omega_0=\Omega_2$, and their negative values $-\Omega_1$, $-\Omega_2$. Explicitly, one finds the values $\Omega_i=\Omega + (-1)^ig$. Only eigenstates presenting these four frequency differences can be connected by the operators $D_{\mu,l}(\Omega_i)$. The states $\ket{1}$ and $\ket{2}$ are not connected by the dipole operator because they contain the same number of excitations.
This becomes more evident by expressing the components of the dipole operators as $D_{\mu, l} (\Omega_i) = d_{\mu,l} L_\mu(\Omega_i)$ in terms of the ladder operators $L_\mu(\Omega_i)$, which in the dressed-state basis take the form:
\begin{equation}
	\label{eq:ladders}
    \begin{split}
         L_\mu(\Omega_1)=&\frac{ \ketbra{2}{3}+(-1)^\mu \ketbra{0}{1}}{\sqrt{2}}=L_ \mu^\dagger(-\Omega_1), \\
         L_\mu(\Omega_2) =&\frac{ \ketbra{0}{2}+(-1)^\mu \ketbra{1}{3}}{\sqrt{2}}=L_ \mu^\dagger(-\Omega_2).
    \end{split}
\end{equation}
Note that $L_\mu(\Omega_i)$ ($L_\mu^\dagger(\Omega_i)$) lowers (excites) only one atom, therefore acting as lowering (raising) operators. 

After expressing the dipole operators, which correspond to the atomic part of $H_{SE}$, in the dressed-state basis, we are now in a position to derive a Markovian master equation for the atomic system. Details of the calculation are found in Appendix \ref{sec:derivemaster}. Here, we write the dynamical equation in the interaction picture with respect to $H_S+H_E$ for the atomic density matrix $\rho_I$ after tracing out the environmental degrees of freedom and making the Born-Markov approximation, leading to
\begin{equation}
    \begin{split}
        \label{eq:LindbladInt}
        &\dot\rho_I =  \sum_{\mu,\mu'=1}^{2}\sum_{\omega,\omega'} e^{i(\omega'-\omega)t}
        \gamma^{\mu,\mu'}(\omega) \times
        \\
        &\Big(L_{\mu'}(\omega) \rho_I L_{\mu}^\dagger (\omega') -           
        \tfrac{1}{2}\{L_{\mu}^\dagger (\omega') L_{\mu'}(\omega), \rho_I  \} \Big),
    \end{split}
\end{equation}
where the frequencies $\omega$ and $\omega'$ can take the four possible values $\pm\Omega_1$ and $\pm \Omega_2$ that were previously discussed. Furthermore, we have introduced the decay rates 
\begin{equation}
		\label{eq:gammas}
    \gamma^{\mu,\mu'}(\omega)=
    \sum_{l,l'=1}^3 
        2\Gamma_{l,l'}^{\mu,\mu'}
        (\omega)d_{\mu,l}^\ast d_{\mu',l'},
\end{equation}
which are given in terms of the spectral correlation tensor that corresponds to the one-sided Fourier transform
\begin{equation}
	\label{eq:GammasInt}
    \Gamma_{l,l'}^{\mu,\mu'}(\omega)=\frac{1}{\hbar^2}
    \int_0^\infty dt' e^{i\omega t'}
    \langle
    E_{\mu,l}(t)E_{\mu',l'}(t-t')
    \rangle.
\end{equation}
Note that, in contrast with Ref. \cite{BreuerPetruccione2002}, we have four indices instead of two. This is because the field couples differently to each atom, as their positions are different; see also Appendix \ref{sec:derivemaster}. Taking the separation between the atoms into account, one is led to the expressions
\begin{align}
	\label{eq:Gammas}
    &\Gamma_{l,l'}^{\mu,\mu'}(\omega)=\delta_{l,l'}(1+n(\omega))\omega^3
    \frac{
    \delta_{\mu,\mu'}+(1-\delta_{\mu,\mu'})f_l(\tau\omega)
    }{6\hbar\epsilon_0\pi c^3},\nonumber
    \\
    %&f_1(\omega)=f_2(\omega)=3\frac{(\chi^2-1)\sin{\chi} +\chi \cos{\chi}}{2\chi^3},
    %\\
    &f_1(\chi)=f_2(\chi)=3\frac{(\chi^2-1)\sin{\chi} +\chi \cos{\chi}}{2\chi^3},
\\
    &f_3(\chi)=3\frac{\sin{\chi} -\chi \cos{\chi}}{\chi^3},\quad \chi=\frac{2R\Omega}{c},\quad
    \tau=\frac{\chi}{\Omega}.
    \nonumber
    \end{align}
Here, we have used the mean photon occupation number 
    \begin{equation}
    	\label{eq:nocupation}
        n(\omega)=\frac{1}{{e^{\hbar \omega/k_B T}-1}},
    \end{equation}
    given in terms of the Boltzmann constant $k_B$ and the temperature $T$. 
   We have also introduced the dimensionless variable $\chi$, which is reminiscent of the Lamb-Dicke parameter as it is given by a distance multiplied by a wave number; although here it parameterizes the separation between the two atoms.
    Once again, one should note that the possible values of the frequencies are $\omega \in\{\pm\Omega_1,\pm\Omega_2\}$. Observe that $\Gamma^{\mu,\mu}_{l,l'}$ (equal superscripts $\mu$) recovers the result for single atoms. In addition, $f_l(\omega)$ vanishes for large values of $\chi$, that is, for large atomic separations or very large transition frequencies. In the opposite limit, when $\chi\to 0$, each $f_l$ tends to unity. In this case, $\Gamma^{\mu,\mu'}_{l,l'}$, with different values of $\mu$, also coincides with the single-atom case.

 \begin{figure}
 	\centering
 	\includegraphics[width=0.35\textwidth]{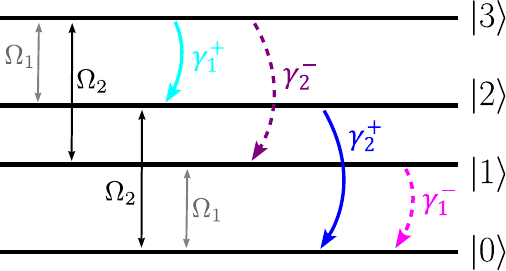}
 	\caption{
 		Energy levels of the dressed states, Eq. \eqref{eq:dressedbasis}, representing the eigenstates of $H_S$. We also present the relevant frequency differences $\Omega_i$, Eq. \eqref{eq:dressedOmegas} or \eqref{eq:OmegasChi}, and the decay rates between dressed states $\gamma^\pm_i$ given in \eqref{eq:gammasi}, for $i\in\{1,2\}$. Note that no decay mechanism occurs between states $\ket{2}$ and $\ket{1}$ as both possess the same number of excitations. The dissipative processes connecting to $\ket{1}$, corresponding to $\gamma_i^-$, are depicted with dashed lines, as they can be greatly suppressed, leading to a transient protection of the maximally entangled state $\ket{1}$.}
 	\label{fig:levels}
 \end{figure}
 \subsection{The dressed state master equation}
  In what follows, we will assume that both dipole moments are equal, $|\vec d_1|=|\vec d_2|\equiv |\vec d|$, and are orthogonal to the line connecting the atoms. This is a reasonable physical assumption, as one could consider two identical atomic species subjected to a constant electric field that aligns their dipoles. Under the rotating-wave approximation (RWA), we only keep terms with the same energy difference where $\omega=\omega'$ in Eq. \eqref{eq:LindbladInt}. Then, transforming back to the Schrödinger picture and expressing the jump operators connecting the dressed states in the form $\ketbra{j}{j'}$ using Eq. \eqref{eq:ladders}, one is led to the dressed-state master equation in Lindblad form
 \begin{align}\label{eq:master}
 	&\dot\rho=\cL\rho=\frac{1}{i\hbar}[H_S,\rho]+   
 	\gamma_1^+(1+ n_1)\cD_{23}\rho+\gamma_1^+  n_1\cD_{32}\rho\nonumber\\
 	&
 	\gamma_2^+(1+ n_2)\cD_{02}\rho+\gamma_2^+  n_2\cD_{20}\rho
 	+
 	\gamma_2^-(1+ n_2)\cD_{13}\rho+\nonumber\\
 	&\gamma_2^-  n_2\cD_{31}\rho+
 	\gamma_1^-(1+ n_1)\cD_{01}\rho+\gamma_1^-  n_1\cD_{10}\rho,
 \end{align}
 with $n_i=n(\Omega_i)$ and with the introduction of the dissipators in the dressed-state basis
 \begin{equation}\label{eq:dissipator}
 	\mathcal{D}_{j'j}\rho=\frac{1}{2}\big(2\ketbra{j'}{j}\rho\ketbra{j}{j'}-\ketbra{j}{j}\rho-\rho\ketbra{j}{j}\big).
 \end{equation}
 Furthermore, the decay rates between pairs of dressed states is given by
\begin{equation}
	\label{eq:gammasi}
          \gamma_i^\pm= \gamma\frac{\Omega_i^3}{\Omega^3} \left[1\pm f_1(\chi\Omega_i/\Omega)\right], \quad
          \gamma= \frac{\Omega^3|{\vec d}|^2}{3 \hbar c^3\pi\epsilon_0 },
\end{equation}
where we have chosen to label them in terms of the $i$-th energy difference, $\Omega_i$, and a sign $-$ ($+$) for the transitions connecting with the one-excitation state $\ket{1}$ ($\ket{2}$), as depicted in Fig.~\ref{fig:levels}. The parameter $\gamma$ is precisely the spontaneous emission decay rate in the single-atom case. We remind the reader that $\Omega$ is the bare atomic transition frequency, and $\Omega_i=\Omega+(-1)^ig$ where $g$ is the dipole-dipole coupling frequency. This latter constant can be explicitly determined from the dipole-dipole interaction energy which, under the previous assumptions, leads to the value
\begin{equation}
	\label{eq:g-dipole-dipole}
    g=\frac{\vec{d_1}\cdot\vec{d_2}}{4\pi\epsilon_0\hbar R^3}=
    \frac{6\gamma}{\chi^3}
    , \quad \chi=
    \frac{2R\Omega}{c}.
\end{equation}
 In terms of $\chi$, the frequency difference between dressed states can be rewritten in the following form
\begin{equation}
	\label{eq:OmegasChi}
\Omega_i=  \Omega\left( 1+(-1)^i
\frac{\gamma}{\Omega}\frac{6}{\chi^3}
\right).
\end{equation}

The master equation in \eqref{eq:master} might look complicated; however, it presents a very interesting and mathematically convenient feature: every off-diagonal element is an eigenvector of the Liouvillian $\cL$. This can be confirmed by letting the dissipator act on any off-diagonal term as $\cD_{j'j}\ketbra{k}{k'}=-\ketbra{k}{k'}(\delta_{j,k}+\delta_{j,k'})/2$ for $k\neq k'$. Note that these elements are also eigenvectors of the commutator part, as each state $\ket{j}$ is an eigenstate of $H_S$. In this sense, the structure of the Liouvillian becomes exceptionally transparent, allowing the diagonalization problem to be treated in a much more compact form.

\subsection{Regime of validity}
We have successfully cast a dynamical equation for two interacting dipoles in terms of three main parameters: the atomic transition frequency $\Omega$, the single-atom decay rate $\gamma$, and the dimensionless parameter $\chi$ that is proportional to the atomic separation $2R$ and $\Omega$. In deriving the dressed-state master equation, we applied the rotating-wave approximation, which requires a clear separation between the frequency differences $\Omega_i$ in Eq. \eqref{eq:OmegasChi}. A fourth parameter also comes into play: the temperature $T$, which fixes the mean photon occupation number and also depends on $\Omega$ as shown in \eqref{eq:nocupation}. For large frequencies, close to the optical regime, the mean photon occupation number is negligible even at room temperature.

However, optical frequencies increase the value of $\chi$, and if it becomes too large ($\chi^3\gg6\gamma/\Omega$), the approximation fails as each $\Omega_i$ tends to $\Omega$. Conversely, if $\chi$ is small enough ($\chi^3\to6\gamma/\Omega$), then $\Omega_1$ approaches zero, leading to a close proximity in the frequencies $\pm\Omega_1$, which would invalidate the RWA and, consequently, the dynamical equation. A relevant quantity is therefore the ratio between the single-atom decay rate $\gamma$ and the atomic frequency $\Omega$. Typically, the latter is several orders of magnitude larger than the former, i.e., $\Omega\gg\gamma$. 
  
\begin{figure*}
\includegraphics[width=.35\textheight]{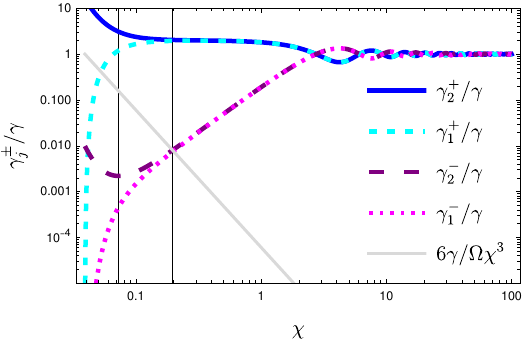}
\includegraphics[width=.35\textheight]{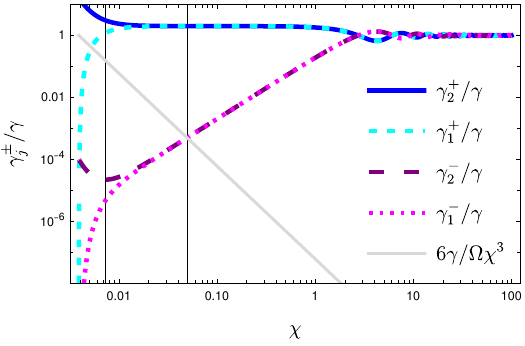}
\caption{\label{fig:decays}
	Decay rates between dressed states, as shown in Fig.~\ref{fig:levels}, for two different values of the ratio $\gamma/\Omega$ between the single-atom spontaneous emission rate $\gamma$ and the atomic frequency $\Omega$. In the left (right) panel, $\gamma/\Omega=10^{-5}$ ($\gamma/\Omega=10^{-8}$). The interval between the vertical lines at $\chi_-=(39\gamma/\Omega)^{1/3}$ and $\chi_+=(30\gamma/\Omega)^{1/5}$ exhibits a clear separation between a fast time scale, governed by $\gamma_i^+$ and involving the dressed state $\ket{2}$, and a transient time scale where $\ket{1}$ is protected, slowly decaying at the rates $\gamma_i^-$. Furthermore, at least within this interval, the rotating-wave approximation remains valid as the eigenvalues of $H_S$ are well separated and do not approach degeneracy. 
} 
\end{figure*}

In Fig.~\ref{fig:decays}, we present, on a logarithmic scale, the four decay rates $\gamma_i^\pm$ normalized by $\gamma$ for two different ratios: $\Omega/\gamma=10^5$ (left panel) and $\Omega/\gamma=10^8$ (right panel). We also include the normalized frequency correction $g/\Omega=6\gamma/(\Omega\chi^3)$ as a solid gray line; the plot starts on the horizontal axis when this value reaches unity, which marks the boundary of the validity regime. For large values of $\chi$, all values of $\gamma_i^\pm$ converge to $\gamma$, marking another boundary outside the range of validity.

In contrast, for intermediate values of $\chi$, one finds the most interesting physical behavior: the two $\gamma_i^-$ rates are several orders of magnitude smaller than the $\gamma_i^+$ rates. This implies that the state $\ket{1}$ has a much smaller coupling to other dressed states in comparison to $\ket{2}$, as can be seen in Fig.~\ref{fig:levels}. The decay of $\ket{1}$ never strictly vanishes; however, its dynamics unfold on a vastly slower time scale. The challenge in satisfying these conditions experimentally is that either the atomic distance $2R$ or the frequency $\Omega$ must be small enough to attain a value of $\chi$ smaller than one. For this reason, frequencies well below the optical regime might be preferable, with the drawback that the thermal occupation number in this regime is no longer negligible at room temperature. In the next section, we will first investigate the case of zero temperature ($n_i=0$) to gain fundamental insight into the features of the system. Following that, we will tackle the finite temperature case and demonstrate that the protection of $\ket{1}$ persists.

We finalize this section with an estimation of a regime of validity in terms of upper and lower bounds for $\chi$.
From Fig.~\ref{fig:decays}, one can note that each $\gamma_i^-$ can be orders of magnitude smaller than any $\gamma_i^+$. This clear temporal separation occurs, at minimum, in the region between the minimum value of $\gamma_2^-/\gamma$ and the point where it crosses the line $6\gamma/(\Omega\chi^3)$. These two bounds, for a given ratio $\gamma/\Omega$, can be obtained by Taylor expanding $f_1(\chi)\approx 1-\chi^2/5$ for small arguments. Then, from Eq. \eqref{eq:gammasi}, one finds that $\gamma_2^-/\gamma\approx(1+g/\Omega)^5\chi^2/5$ with $g=6\gamma/\chi^3$ given in \eqref{eq:g-dipole-dipole}. A critical point analysis of this function yields a minimum value for $\gamma_2^-/\gamma$ at $\chi_-=(39\gamma/\Omega)^{1/3}$. For the crossing with the gray line, one notes that $g/\Omega$ decreases and therefore equating $\chi^2/5=g/\Omega$ leads to the value $\chi_+=(30\gamma/\Omega)^{1/5}$.

%%%%%%%%%%%%%%%%%%%%%%%%%%%%%%%%%%%%%%%%%%%%%%%%%%%%%%%%%%%%%%%%%%%%%%%%%%%%%%%%%%%%%%%%%%%%%%%%%%%%%%
\section{Zero temperature bath}  \label{sec:zerotemp}
\subsection{Solution to the eigenvalue problem}
If the atomic transition frequency is large enough, for instance in the optical regime, the mean photon numbers $n_1$ and $n_2$ are negligible and the Lindbladian in Eq. \eqref{eq:master} presents only losses. In this case, the problem is tractable in the sense that the full eigensystem can be obtained in closed form \cite{Torres2014}. In order to do this, it is more convenient to rewrite the Lindbladian
in Eq. \eqref{eq:master} as 
\begin{equation}
\cL=\cK+\cA,\quad \cK\rho=-i(K\rho-\rho K^\dagger)/\hbar.
\end{equation}
where we have introduced the jump superoperator $\cA$ that will be shortly specified, and the superoperator $\cK$ given in terms
of the effective non-Hermitian Hamiltonian $K$ which is diagonal in the eigenbasis of $H_S$, namely
\begin{equation}
    K = \sum_{j=0}^3 \hbar\zeta_j \ketbra{j}{j}.
\end{equation}
We have introduced the complex-valued eigenfrequencies that are simply obtained from the eigenfrequencies of $H_S$ plus an imaginary part coming from the dissipative part that take the following form
\begin{align}
\zeta_0&=0, \quad \zeta_1=\Omega - g-i \frac{\gamma_1^-}{2}, \nonumber\\
\zeta_2&= \Omega + g-i \frac{\gamma_2^+}{2},\quad
\zeta_3=2 \Omega -i\frac{ \gamma_1^++ \gamma_2^-}{2}.
\end{align}
 Note that $K$ is not only diagonal in the eigenbasis of $H_S$, it is in fact a normal operator as $[K,K^\dagger]=0$. 
 In contrast with phenomenological master equations that typically require left and right eigenstates of $K$ \cite{Torres2014}, here these two are the same. This fact greatly simplifies the diagonalization  of the Liouvillian $\mathcal{L}$. 
 
As for the jump operator $\cA$, it is not hard to realize that its action on an arbitrary density matrix is given by
\begin{align}\label{eq:jump}
	\mathcal{A}\rho =&  \bra3\rho\ket3 \left(
	\gamma_1^+ \ketbra{2}{2} +
    \gamma_2^-  \ketbra{1}{1}\right) \nonumber\\
    &+
	(\gamma_1^- \bra1\rho\ket1+\gamma_2^+\bra2\rho\ket2) \ketbra{0}{0} 
\end{align}
with the matrix elements  $\rho_{jj}=\bra{j}\rho\ket{j}$ of the density operator. Note that only the diagonal elements are coupled by $\cA$. In fact, this superoperator kills all off-diagonal elements. Therefore, every off-diagonal element is an eigenvector of $\cL$, which 
can be corroborated by direct application as $\cL\ketbra{j}{j'}=\cK\ketbra{j}{j'}$ whenever $j\neq j'$ with
\begin{equation}
\cK\ketbra{j}{j'}=-i(\zeta_j-\zeta_{j'}^\ast)\ketbra{j}{j'}.
\end{equation}
The former equation also applies to the diagonal elements, when $j=j'$, and together with \eqref{eq:jump} one is able to find the action
of $\cL$ as
\begin{equation}
	\begin{split}
		\mathcal{L}\ketbra{3}{3} & = 
		\gamma_1^+ \ketbra{2}{2} + \gamma_2^- \ketbra{1}{1} -(\gamma_2^-+\gamma_1^+) \ketbra{3}{3} ,\\
		\mathcal{L}\ketbra{2}{2} & =  -\gamma_2^+ \ketbra{2}{2} +
		\gamma_2^+\ketbra{0}{0}, \\
		\mathcal{L}\ketbra{1}{1} & =  -\gamma_1^- \ketbra{1}{1} +
		\gamma_1^-\ketbra{0}{0}, \quad
		\mathcal{L}\ketbra{0}{0}  =  0.
	\end{split}
\end{equation}
From these equations we immediately identify the stationary state
$\hat{\rho}_0 = \ketbra{0}{0}$ and the linear combinations that describe the
relaxing modes.
Analogous expressions can be found for the adjoint Liouvillian $\cL^\dagger$, by simply noting that for a superoperator 
$\mathcal O \rho=A\rho B$, with $A$ and $B$ being operators, the adjoint superoperator acts as $\mathcal O^\dagger \rho=A^\dagger \rho B^\dagger$. 
By realizing that the Liouvillian is triangular and using the technique in Ref. \cite{Torres2014}, one can solve the dual eigenvalue equation
\begin{equation} 
		\mathcal{L}\hat{\rho}_\lambda = \lambda \hat{\rho}_\lambda,\quad 
	\mathcal{L}^\dagger \check{\rho}_\lambda = \lambda^\ast \check{\rho}_\lambda
\end{equation} 
with the complex eigenvalue $\lambda$ that labels the right and left eigenvectors, $\hat\rho_\lambda$ and $\check \rho_\lambda$,  which form an orthonormal basis 
with the aid of the Hilbert-Schmidt inner product 
$ {\rm Tr}\{\check{\rho}_\lambda^\dagger \hat{\rho}_{\lambda'} \} = \delta_{\lambda,\lambda'}$. Table \ref{tab:eigensystem} summarizes the eigenvalues and eigenvectors of the Liouvillian superoperator $\mathcal{L}$.

\begin{table}[h]
\begin{tabular}{c|c|c}

$\lambda$     & $\hat{\rho}_\lambda$               & $\check{\rho}_\lambda^\dagger$                   \\ \hline
$0$                      & $\ket{0}\bra{0}$                   & $\mathbbm{1}$         \\ \hline
$-\gamma_1^-$            & $\ket{1}\bra{1}-\ketbra{0}{0}$     & $\frac{\gamma_1^+}{\gamma_1^++\gamma_2^--\gamma_1^-} \ketbra{3}{3}+\ketbra{1}{1}$                     \\ \hline
$- \gamma_2^+$            & $\ketbra{2}{2}-\ketbra{0}{0}$     & $\frac{\gamma_2^-}{\gamma_1^++\gamma_2^--\gamma_1^-}\ketbra{3}{3}+\ketbra{2}{2}$        \\ \hline
$- \gamma_1^+-\gamma_2^-$ & $\displaystyle\sum_{j=0}^3 d_j \ket{j} \bra{j}$ & $\ket{3}\bra{3}$                     \\ \hline
$i\zeta_{j'}^\ast-i\zeta_{j}$ 
& $\ketbra{j}{j'}$                     & $\ketbra{j}{j'}$                      
\end{tabular}
 \caption{\label{tab:eigensystem}Eigenvalues $\lambda$ and eigenvectors $\hat\rho_\lambda$ of $\mathcal{L}$, and eigenvectors $\check\rho_\lambda$ of the adjoint operator $\cL^\dagger$. We have used the coefficients $d_3=1$,
 	$d_2 = \frac{\gamma_2^-}{\gamma_2^+ - \gamma_1^+ - \gamma_2^-}$,
	$d_1 = \frac{\gamma_1^+}{\gamma_1^- - \gamma_1^+ - \gamma_2^-}$, and
	$d_0 = \frac{\gamma_2^+\gamma_2^- + \gamma_1^-\gamma_1^+ - \gamma_1^-\gamma_2^+}
	{(\gamma_2^+ - \gamma_1^+ - \gamma_2^-)(\gamma_1^- - \gamma_1^+ - \gamma_2^-)}$.
 }
\end{table}

\subsection{Time evolution}\label{sec:timeevolution}
Having solved the eigenvalue problem, it is now possible to study the dynamical features of the system. In particular, due to the simplicity of the solutions, it is possible to express the time-dependent density matrix in a compact form, even for an arbitrary initial condition such as
\begin{equation}  
	\rho(t=0) = \sum_{j,j'} P_{jj'} \ketbra{j}{j'},\quad \rho_{jj'}(0)=P_{jj'}.
\end{equation}  
The coefficients $P_{jj'}$ represent the matrix elements of the initial density matrix $\rho_0$ in the energy (dressed-state) basis.
Exploiting the completeness of the eigenbasis of $\cL$, the density operator at arbitrary times can be evaluated as
\begin{equation}\label{eq:evolucionrho} 
\rho(t)=\sum \operatorname{Tr} \left\{\check{\rho_\lambda}^\dagger\rho_0 \right\}e^{\lambda t}\hat{\rho}_\lambda, 
\end{equation}
where the sum runs over the eigenvalues $\lambda$ of the Liouvillian superoperator $\cL$, with corresponding right eigenvectors $\hat{\rho}_\lambda$ and left eigenvectors $\check{\rho}_\lambda^\dagger$.
After evaluating the traces, the resulting density matrix at time $t$ can be written as
\begin{align}
	\label{eq:solutionzero}
    \rho_{11}(t) & =
    P_{33}\gamma_2^-\frac{e^{-\gamma_1^- t}-e^{-(\gamma_1^++\gamma_2^-)t}}{\gamma_1^+ + \gamma_2^- - \gamma_1^-}
     + P_{11}e^{-\gamma_1^- t},    \\
    \rho_{22}(t) & = 
     P_{33}\gamma_1^+\frac{e^{-\gamma_2^+ t}-e^{-(\gamma_1^+ + \gamma_2^-)t }}{\gamma_1^+ + \gamma_2^- - \gamma_2^+}
     + P_{22}e^{-\gamma_2^+ t},    \\
    \rho_{33}(t) & = P_{33}\, e^{-(\gamma_1^+ + \gamma_2^-) t},\quad \rho_{jk}(t)= P_{jk}e^{i(\zeta_j-\zeta_k)t}.
\end{align}
The time-dependent probability of the ground state, $\rho_{00}(t)$, is obtained from the unit trace condition of $\rho$. Note that
this is the only surviving term in the limit $t\to\infty$, meaning that $\ket{0}$ is the only steady state. It is, however, interesting to note that the slowest decay rate, $\gamma_1^-$, maintains the state $\ket{1}$ for longer times. In the limit of vanishing $\gamma_1^-$, two  steady states are present, namely $\ket0$ and $\ket1$. Given the fact that $\ket 1$ is preserved, at least for a transient time, it is relevant to study the degree of entanglement as a function of time. 

\subsection{Entanglement}
As we are working with mixed states, we will employ the concurrence \cite{Wootters98}, which is a widely used measure of entanglement for two-level quantum systems (qubits). It ranges from 0 for separable states to 1 for maximally entangled states, providing a clear quantification of this type of quantum correlations. Its operational definition is given in Appendix \ref{sec:entanglement}. In our setting, to analyze entanglement properties in the system, we consider a pure initial state of the form
\begin{equation}
	\frac{c_1\ket{1}+c_2\ket2}{\sqrt2}=\frac{(c_1+c_2)\ket{eg}+(c_2-c_1)\ket{ge}}{2}.
\end{equation}
We have chosen  such a condition, as it includes separable states with just one excited atom ($c_2=\pm c_1=1/\sqrt2$), or completely entangled states when $c_1=1$ or when $c_2=1$. The former case might be the more relevant as it is the most feasible to prepare experimentally. Using the solutions given in Eq. \eqref{eq:solutionzero}, one can obtain the nonzero entries of the time-dependent density matrix as
\begin{align}
    \rho_{22}(t) &=|c_2|^2e^{-\gamma_2^+ t},
    \quad \rho_{11}=|c_1|^2 e^{-\gamma_1^- t},\nonumber\\
    \rho_{12}(t)&=c_1c_2^\ast e^{2ig t-(\gamma_1^-+\gamma_2^+)t/2}
\end{align}
As shown in Appendix \ref{sec:entanglement}, for states of this type one can evaluate the concurrence in the simple form
\begin{equation}
	\label{eq:xsimpleconc}
	C=\sqrt{(\rho_{22}-\rho_{11})^2-(\rho_{12}-\rho_{21})^2
		%4({\rm Im}[\rho_{21}])^2
		}.
\end{equation}
%\begin{equation*}
%	C=\sqrt{(|c_2|^2 e^{-\gamma_2^+ t}-|c_1|^2e^{-\gamma_1^- t})^2+4 e^{-\gamma_2^+t-\gamma_1^-t}({\rm Im}[c_1c_2^\ast e^{2igt}])^2}
%	\end{equation*}
In the regime of interest, $\gamma_2^+\gg\gamma_1^-$, and in the long-time limit, the concurrence takes the 
simple expression $|c_1|^2e^{-\gamma_1^- t}$.
In the particular case of an excited atom, the result greatly simplifies to
\begin{equation}
	\label{eq:conczero}
	C=\tfrac{1}{2}\sqrt{(e^{-\gamma_2^+ t}-e^{-\gamma_1^- t})^2+4 e^{-(\gamma_2^++\gamma_1^-)t}\sin^2(2gt)}
\end{equation}
with the long-time limit behavior $C\to e^{-\gamma_1^-t}/2$. Of course, the effect is transient and eventually the state
reaches the ground state. However, there is a time scale  $1/\gamma_2^+\ll t \ll 1/\gamma^-_1$ during which the state persistently remains in the incoherent superposition $(\ketbra{1}+\ketbra{2})/2$, presenting a moderate degree of entanglement. It is remarkable that the dissipative dynamics described by the dressed-state master equation is able to produce a lasting entanglement from a completely separable pure state.

%In this regime, the system is effectively behaving as if maximally entangled state $\ket1$ was a second steady state arising from dissipator
%    \begin{equation}\label{eq.D.final}
%	\begin{split}
%		\mathcal D[\rho] &=     
%		2\gamma(\omega_1)\Big( \ketbra{2}{3} \rho \ketbra{3}{2} -  \frac{1}{2}\{\ketbra{3}{3}, \rho  \}  \Big) \\
%		& +2\gamma(\omega_2) \Big( \ketbra{0}{2} \rho \ketbra{2}{0} -  \frac{1}{2}\{\ketbra{2}{2}, \rho  \}  \Big) \\
%	\end{split}
%\end{equation} 
%%%%%%%%%%%%%%%%%%%%%%%%%%%%%%%%%%%%%%%%%%%%%%%%%%%%%%%%%%%%%%%%%%%%%%%%%%%%%%%%%%%%%%%%%%%%%%%%%%%%%%
\section{Finite temperature}  \label{sec:nonzerotemp}
We devote this section to study the finite temperature case. As
previously discussed, this situation is relevant in a system presenting atomic transition frequency well below the optical regime. In such a case, the thermal photons cannot be neglected. In contrast, 
the  atomic separation does not need to be too small to achieve relevant values of the parameter $\chi$, therefore, leading to a more feasible realization. 
In this more general case, the off-diagonal elements are still eigenvectors of $\cL$ and their temporal behavior presents damped oscillations until the steady state is reached. Although the eigenvectors for the diagonal elements are, in principle, also possible to calculate exactly, this relies on the roots of third order polynomial.  
In order to simplify the problem, we take advantage of the
 transient time where the entangled state $\ket1$ is decoupled from the dynamics, and one can approximate $\gamma_1^-$ and $\gamma_2^-$ to zero. As before, the Liouvillian couples only the diagonal elements, but now it can be restricted to the basis
$\{\ketbra{0},\ketbra{2},\ketbra{3}\}$, where it takes the matrix representation
\begin{equation}
	\label{eq:Ltemp}
	L=
	\begin{pmatrix}
		-\gamma_2^+n_2 &\gamma_2^+(n_2+1)& 0\\
		\gamma_2^+n_2 &-\gamma_2^+(n_2+1)-\gamma_1^+n_1& \gamma_1^+(n_1+1)\\
		0 &\gamma_1^+n_1& -\gamma_1^+(n_1+1)
	\end{pmatrix},
\end{equation}
with each matrix element obtained using the relation $L_{jj'}=\bra{j'}(\cL\ketbra j)\ket{j'}$. Eigenvalues and eigenvectors can be obtained: one is zero, and the others are roots of a second degree polynomial. In this case, however, we will restrict our analysis  to the steady state of this matrix, the one corresponding to the zero eigenvalue, whose nonzero matrix elements are found to be
\begin{align}
	\rho_{00}^{\rm st}&=(n_1+1)(n_2+1)/s,\quad 
	\rho_{22}^{\rm st}=(n_1+1)n_2/s
	\nonumber\\
	\rho_{33}^{\rm st}&=n_1n_2/s, \quad
	s=1+n_1+2n_2+3n_1n_2.
\end{align}
This state, together with $\ketbra{1}$, are now the two steady states of the system
in an approximate way before a time is reached where the decay produced by $\gamma_i^-$ becomes noticeable. Note, however, from Fig.~\ref{fig:decays} that these constants are orders of magnitude smaller than their counterparts $\gamma_i^+$. 

\begin{figure}
	    \centering
	
	\includegraphics[width=.47\textwidth]{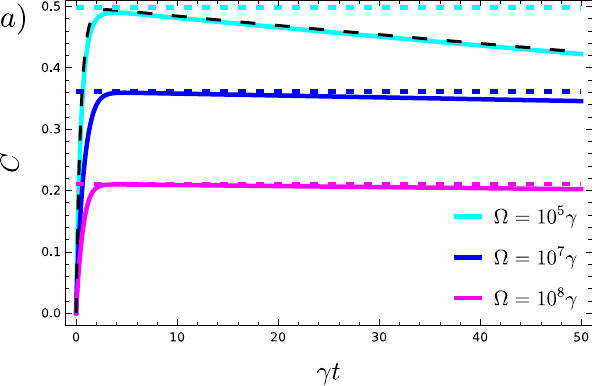}
   \vspace{7pt}
   
		\includegraphics[width=.47\textwidth]{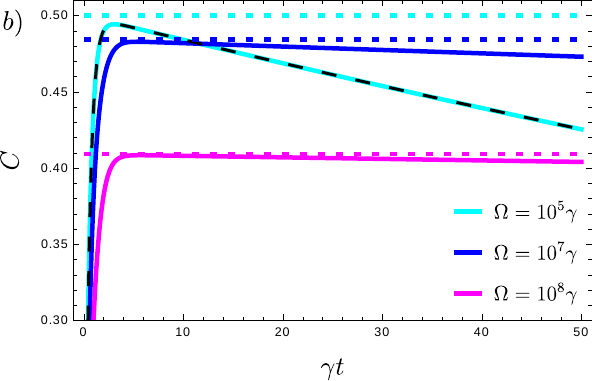}
	\caption{\label{fig:concurrence}
Concurrence as a function of time, presented in solid curves, for the initial state $\ket{eg}$ and different values of the ratio between the atomic frequency and the free atomic decay rate
$\Omega/\gamma\in \{10^{5}, 10^{7}, 10^{8}\}$. 
In colored-dashed lines we have included the prediction of the concurrence given 
in Eq. \eqref{eq:conctemp}. We have chosen an interatomic distance of $1\mu $m in $a)$, and of $0.5\mu $m in $b)$. This also fixes the mean photon numbers for the three mentioned cases. Their values for $a)$ approximately  $n_1\in \{0.001,0.11,0.34\}$ and $n_2\in\{0.002,0.12,0.34\}$, whereas for $b)$ one has
$n_1\in \{10^{-6},0.01,0.07\}$ and $n_2\in\{10^{-6},0.01,0.07\}$. The dashed-black curve corresponds to the approximation for zero mean photon number given in Eq. 
\eqref{eq:conczero}. We have also considered temperature of $T=300K$.
%In the case $\Omega=10^4\gamma$, one achieves small occupation numbers and the prediction for zero temperature in Eq. \eqref{eq:conczero} and shown with a dashed black line is an excellent prediction.
}
\end{figure}

With this result, it is now simple to obtain the value of the
concurrence in the steady state (or transient steady state).
The state is diagonal in the dressed state basis, but is an $X$-state in the computational basis. Therefore, using Eqs. \eqref{eq:concx} and 
\eqref{eq:rhotrans} one finds the simple expression
\begin{align}
	\label{eq:conctemp}
	C_{\rm st}&=
	{\rm Max}\left\{0,
	\left|n_{12}\tilde P_{11}-P_{11}\right|-2\tilde P_{11} \sqrt{n_{12}n_{21}},
	\right\}\nonumber\\
	n_{ij}&=(n_i+1)n_j/s, \quad \tilde P_{11} =1-P_{11}.
\end{align}
In Fig.~\ref{fig:concurrence} we have numerically calculated the concurrence for the initial state $\ket{eg}$, that is, a separable state with one excited atom and the other in the ground state.  We have considered three different ratios between the atomic frequency $\Omega$, and the free spontaneous decay $\gamma$ given in the legend of the plot. We have chosen the value of 
$\chi$ as the mean value between both $\chi_\pm$.
The transient steady state value of the concurrence, Eq. \eqref{eq:conctemp}, is plotted in dashed line of the corresponding color. Note that there is no initial entanglement. It is the interaction between the atoms that creates this correlation and the combined decay mechanism 
between dressed states is what maintains the effect. One can note that a  higher degree of entanglement, in these examples, is achieved for $\gamma/\Omega=10^{-5}$. However, its decay is faster than the other two cases. For a smaller interatomic separation, as in $b)$, the amount of entanglement in the steady state increases. This can be understood as the values of the atomic frequencies are bounded by 
$(30\gamma/\Omega)^{1/5}<2R\Omega_i/c<(39\gamma/\Omega)^{1/3}$. This means that  low values of $\gamma/\Omega$ also imply lower values of the 
frequency, leading to larger values of the mean occupation number, Eq.~\eqref{eq:nocupation}. In order to increase this value, smaller interatomic distances are required, as 
shown in Fig.~\ref{fig:concurrence} $b)$ where this value is of $0.5\mu$m compared to $1 \mu$m in $a)$. This result also demonstrates that small interatomic distances, leading to strong atomic coupling, are required in order to magnify the effect. Although we have chosen
interatomic distances of the order of one micron, smaller values might also be considered \cite{Du2023,YanesThomas2026}.

The ratio between the decay rates can also be estimated from Eq. \eqref{eq:gammasi} and \eqref{eq:g-dipole-dipole}, leading to 
$\gamma_i^+/\gamma_i^-\approx 10/\chi^2$, where $\gamma_i^+\approx 2\gamma$ in the regime of validity of our master equation, as can be corroborated in Fig.~\ref{fig:decays}. The smaller the value of $\chi$, reminiscent of the Lamb-Dicke parameter, the larger the ratio between the decay rates. This also confirms the necessity of small interatomic distances, for large values of transition frequencies, in order to see the effect. Another parameter that can also enter into play is the temperature. In our calculations we have chosen $T=300,$K, similar to ambient temperature. A smaller temperature would lead to smaller occupation numbers and also a higher transient entanglement in the the blue and magenta curve in Fig.~\ref{fig:concurrence}.

%%%%%%%%%%%%%%%%%%%%%%%%%%%%%%%%%%%%%%%%%%%%%%%%%%%%%%%%%%%%%%%%%%%%%%%%%%%%%%%%%%%%%%%%%%%%%%%%%%%%%%
\section{Comparison with the phenomenological master equation}  
\label{sec:Phenomenological}
To finalize this work, let us compare our system with the so-called phenomenological master equation. In such a case, the interaction between the atoms is considered after deriving the Lindblad master equation. This is a 
common approach in the literature to describe losses in quantum systems, where the separate equations are  added to the interaction in order to arrive at the phenomenological master equation 
\begin{equation} \label{eq:feno22}
	\Dot{\rho} =
	\dfrac{1}{i \hbar} [H_S, \rho] +
	\sum_j \dfrac{\gamma_j}{2} \left(2\sigma_j^- \rho \sigma_j^+ - \lbrace \sigma_j^+ \sigma_j^-, \rho \rbrace \right).
\end{equation}
This system is also solvable, due to the small dimensionality \cite{delvalle2010}. For instance, the time-dependent solution of the phenomenological master equation for diagonal initial states and a zero temperature bath is given by
\begin{align}
%\rho_{00}&=1 + P_{33}e^{-\gamma t}(e^{-\gamma t} - 2) - (P_{11} + P_{22})e^{-\gamma t}\\
\rho_{11}&= e^{-\gamma t}(P_{11} + P_{33} - P_{33}e^{-\gamma t})\nonumber\\
\rho_{22}&=e^{-\gamma t}(P_{22} + P_{33} - P_{33}e^{-\gamma t})\nonumber\\
\rho_{33}&=P_{33}e^{-2\gamma t}
\end{align}
The solution in Eq. \eqref{eq:solutionzero} reduces to this expression in the limit of large values of $\chi$. This can be noted from Fig.~\ref{fig:decays}, where all decay rates tend to one as $\chi$ increases. One can also note this analytically from \eqref{eq:gammas}, as 
$f_i(\chi)$ tends to zero for increasing $\chi$. This result is satisfactory, as both models coincide for weakly interacting atoms in their diagonal elements.

The situation changes significantly for the off-diagonal terms, which in this case are not eigenvectors of $\cL$. In fact, here many of them are coupled so that the eigenvectors have to be obtained in a recursive form.
Here, it is not possible to find an agreement between both master equations. The reason for this is that the rotating-wave approximation that we performed in the derivation is not valid here. Before we assumed no degeneracy between eigenenergies, but in the absence of interaction there are degenerate eigenstates. Therefore, the dressed state master equation does not reduce completely to the phenomenological description and both approaches have to be treated in different regimes. 

%%%%%%%%%%%%%%%%%%%%%%%%%%%%%%%%%%%%%%%%%%%%%%%%%%%%%%%%%%%%%%%%%%%%%%%%%%%%%%%%%%%%%%%%%%%%%%%%%%%%%%
\section{Conclusions}\label{sec:conclusions}

We have derived a dressed-state master equation in Lindblad form for two two-level atoms presenting a strong dipole-dipole coupling. 
The interatomic distance plays a key role in the derivation, as the electromagnetic field differs at each atomic position. 
We have identified the necessary conditions and parameter regime where the proposed formalism is valid. Extremely small and excessively large interatomic distances lead to degeneracies in the central system Hamiltonian, which would invalidate the performed rotating-wave approximation.
Consequently, the presented approach does not reduce to the phenomenological master equation in the weak coupling limit, but rather complements the description in the strong coupling regime. 
 In the regime of validity, we find two separate time scales. A fast scale, where the system relaxes into two states: the global steady state and a transient maximally entangled state. Subsequently, on a longer time scale, the transient state slowly decays into the global steady state of the system. This finding corroborates the capability of the strong coupling formalism  to generate and preserve, at least transiently, entanglement.
  Furthermore, we have shown that the eigensystem of the Liouvillian can be calculated in compact form, with each off-diagonal element as an eigenvector. This feature facilitates the analytical evaluation of the time-dependent solutions, leading to compact expressions for the system properties such as the concurrence. Finally, this dressed-state approach provides a more tractable description of the system by expressing solutions directly in the dressed-state basis, a feature we expect to be helpful in describing more elaborate open quantum systems with strong internal coupling.

%%%%%%%%%%%%%%%%%%%%%%%%%%%%%%%%%%%%%%%%%%%%%%%%%%%%%%%%%%%%%%%%%%%%%%%%%%%%%%%%%%%%%%%%%%%%%%%%%%%%%%
\begin{acknowledgements}
This work was supported by CONAHCYT (SECIHTI-Mexico) Research Grant
CF-2023-I-1751. 
\end{acknowledgements}

\appendix
\section{Details of the derivation of the dressed state master equation}
\label{sec:derivemaster}
In this appendix, we present some of the equations that are needed to complete the derivation of the dressed state master equation.  After integrating the von Neumann equation in the interaction picture with respect to the free Hamiltonian $H_0=H_S+H_E$, taking partial trace over the field degrees of freedom, and assuming the Born-Markov approximation, one is led to the Markovian quantum master equation \cite{BreuerPetruccione2002},
\begin{equation*}
	\dot\rho_I(t)=-\frac{1}{\hbar^2}\int_0^\infty dt' {\rm Tr}_E[H_I(t),H_I(t-t'),\rho_I(t)\otimes \rho_E].
\end{equation*}
We have introduced, in the interaction picture, the atomic density operator $\rho_I(t)$, and the interaction Hamiltonian $H_{I}(t)=U_0^\dagger(t)H_{SE}U_0(t)$ with $U_0(t)=e^{-iH_0t\hbar}$. Furthermore, we have considered
the environment density operator $\rho_E$ as a thermal state and therefore diagonal in the eigenbasis of $H_E$. To arrive at Eq. \eqref{eq:LindbladInt}, it is necessary 
to write the atomic operators in $H_I$ in terms of the ladder operators in \eqref{eq:ladders}.
In taking the partial trace over the field degrees of freedom, the following correlations  for the electromagnetic field operators are needed, 
\begin{equation}
    \begin{split}
        \langle a_{\vec k,\lambda}a^\dagger_{\vec k',\lambda'} \rangle & = \delta_{\vec k \vec k'}\delta_{\lambda \lambda'}(1+n(\omega_k)) \\
        \langle a^\dagger_{\vec k,\lambda}a_{\vec k',\lambda'} \rangle & = \delta_{\vec k \vec k'}\delta_{\lambda \lambda'}n(\omega_k) \\
        \langle a^\dagger_{\vec k,\lambda}a^\dagger_{\vec k',\lambda'} \rangle & = \langle a_{\vec k,\lambda}a_{\vec k',\lambda'} \rangle = 0
    \end{split}
\end{equation}
where $n(\omega_k)=({e^{\beta \hbar \omega_k}-1})^{-1}$ is the thermal photon number. In the optical regime, with transition frequencies $\omega_k$ in the hundreds of terahertz, the occupation number is negligible at room temperature, i.e., $n(\omega_k) \approx 0$, ($ \langle a_{\vec k,\lambda}a^\dagger_{\vec k,\lambda} \rangle\approx1$). 

The explicit form of the spectral correlation tensor is found by evaluating the field correlators. Applying the rotating wave approximation (RWA)  yields:
\begin{equation} 
     \begin{split}
         \Gamma_{l,l'}^{\mu,\mu'}(\omega) = &\frac{1}{\hbar^2}
     \sum_{\vec k}\sum_{\lambda}
  \frac{\hbar \omega_k}{2 \epsilon_0 V}
  e_{\vec k,\lambda}^{l}e_{\vec k,\lambda}^{l'} 
  \times\\
  & \int_0^\infty dt'
  e^{-i(\omega_{k'}-\omega)t'}e^{i[(-1)^{\mu'}-(-1)^\mu ]\vec k \cdot \vec R}.
     \end{split}
\end{equation}
The electromagnetic field polarization vector is denoted by the symbol $\hat{e}_{\vec k,\lambda}$ and obeys the following relations
\begin{equation}
\begin{aligned}\label{eq.vecpolaris}
\vec{k} \cdot \hat{e}_{\vec k,\lambda} & =0,\quad \hat{e}_{\vec k,\lambda}\cdot\hat{e}_{\vec k,\lambda'}  =\delta_{\lambda,\lambda^{\prime}} \\
\sum_{\lambda=1,2} e_{\vec k,\lambda}^l e_{\vec k,\lambda}^{l'} & =\delta_{l,l'}-\frac{k_l k_{l'}}{|\vec{k}|^2}, \quad l, l'=1,2,3 .
\end{aligned}
\end{equation}
The sum over wavevectors is converted to a continuous integral
\begin{equation}
\frac{1}{V} \sum_{\vec{k}} \longrightarrow \int \frac{d^3 k}{(2 \pi)^3}=\frac{1}{(2 \pi)^3 c^3} \int_0^{\infty} d \omega_k \omega_k^2 \int d \Omega,
\end{equation}
where $\omega_k=ck$, and $d\Omega$ denotes integration over solid angle with $\vec k=k(\sin\theta\cos\phi,\sin\theta\sin\phi,\cos\theta)$.
Taking into account the relations of the expression \eqref{eq.vecpolaris}, the angular part involves the integrals
\begin{equation*}
     \int d \Omega \left(\delta_{l,l'}-\frac{k_l k_{l'}}{k^2}\right)
     %e^{2 i  \omega_k R \cos{\theta} /c[(-1)^\nu -(-1)^\mu ]} 
     e^{i \chi_k (1-\delta_{\mu,\mu'}) (-1)^\mu \cos\theta} 
     =\delta_{l,l'} I^{\mu,\mu'}_{l} (\omega_k),
 \end{equation*}
with $\chi_k=2R\omega_k/c$. Evaluating these integrals leads to $I_{l}^{\mu,\mu}=8\pi/3$, for $\mu=\mu'$, that coincides with the single-atom case, whereas for different values of $\mu$ and $\mu'$ one obtains
 \begin{align}
  & I^{\mu,\mu}_i(\omega_k)=\pi\int_{-1}^1(1+u^2)e^{i\chi_k u}du=\frac{8\pi}{3}f_i(\chi_k),\nonumber\\
   &I_3^{\mu,\mu}(\omega_k)=2\pi\int_{-1}^1(1-u^2)e^{i\chi_k u}du=\frac{8\pi}{3}f_3(\chi_k),
 \end{align}
 with $i\in\{1,2\}$ and with the functions $f_j(\chi)$ given in \eqref{eq:Gammas}.
One can arrive at Eq. \eqref{eq:Gammas} after performing the integral over the radial part, considering that the time integral 
leads to $\int_0^\infty dt' e^{-i \omega t'}=\pi\delta(\omega)-i {\rm P}\frac{1}{\omega}$ in terms of delta function and the Cauchy principal value. We omit the imaginary part that leads to a correction of the eigenenergies of the central system \cite{BreuerPetruccione2002}. The delta function contribution leads to a selection of frequencies $\omega_k\to\omega$ with $\omega$ taking the possible values $\pm\Omega_1$ and $\pm\Omega_2$, i.e., the possible frequency difference between dressed states as depicted in Fig.~\ref{fig:levels}.

\section{Two-qubit entanglement}
\label{sec:entanglement}
Entanglement for a general two-qubit state $\rho$ can be evaluated using the concurrence \cite{Wootters98} defined as
\begin{equation}
    C\equiv C(\rho) = \max\{0, \Lambda_1 - \Lambda_2 - \Lambda_3 - \Lambda_4\},
\end{equation}
where $\Lambda_i$ are the square roots of the eigenvalues of the positive matrix $\rho\tilde{\rho}$, where $\tilde{\rho} = \sigma_y^{\otimes 2}\rho^*\sigma_y^{\otimes 2}$ with $\rho^*$ the complex conjugate of $\rho$ in the computational basis
\[\ket{\phi_0}=\ket{gg},\quad \ket{\phi_1}=\ket{ge},\quad \ket{\phi_2}=\ket{eg},\quad \ket{\phi_3}=\ket{ee}.\]
Using this basis, we will only consider $X$ states, where the only non-vanishing elements, $\varrho_{jk}=\bra{\phi_j}\rho\ket{\phi_k}$, of $\rho$ are the diagonal ones, and $\varrho_{03}$, $\varrho_{30}$, $\varrho_{12}$ and $\varrho_{21}$. We have introduced the notation $\varrho_{jk}$ to distinguish the matrix elements in the computational basis from $\rho_{jk}$ in the dressed-state basis in \eqref{eq:dressedbasis}.
For $X$ states, the concurrence takes the simple form
\begin{equation}	
	\label{eq:concx}
	C=2{\rm max}\left\{0,|\varrho_{03}|-\sqrt{\varrho_{22}\varrho_{11}},|\varrho_{12}|-\sqrt{\varrho_{00}\varrho_{33}}\right\}.
\end{equation}
Note that the states labeled by $0$ and $3$ coincide in both bases, and therefore $\rho_{00}=\varrho_{00}$, and $\varrho_{33}=\rho_{33}$. In the other two cases, $j,k\in\{1,2\}$, one can find that 
\begin{equation}
	\label{eq:rhotrans}
\varrho_{jk}=\frac{\rho_{22}+(-1)^{j+k}\rho_{11}+(-1)^j\rho_{12}+(-1)^k\rho_{21}}{2}.\,
\end{equation}
For vanishing values of $\rho_{03}=\varrho_{03}$, and $\rho_{33}$, one has the simple expression given in Eq. \eqref{eq:xsimpleconc} 
that has already been written using the matrix elements in the basis in \eqref{eq:dressedbasis}.

%\nocite{*}

\end{document}